\documentclass[a4paper,11pt,reqno]{amsart}  
\usepackage[latin1]{inputenc}  
\usepackage{enumerate}  
\usepackage{amsfonts}  
\usepackage{amsmath}  
\usepackage{amssymb}  
\usepackage{hyperref}  
  
\newcommand{\RR}{{\mathbb R}}  
\newcommand{\CC}{{\mathbb C}}  
\newcommand{\N}{{\mathbb N}}  
\newcommand{\NN}{{\mathbb N}}

\newcommand{\hil}{\mathcal H}  
\newcommand{\eL}{\mbox{L}}

\DeclareMathOperator{\supp}{supp}

\newtheorem{theorem}{Theorem}[section]  
\newtheorem{coro}[theorem]{Corollary}  
\newtheorem{definition}[theorem]{Definition}  
\newtheorem{lemma}[theorem]{Lemma}  
\newtheorem{prop}[theorem]{Proposition}  
\newtheorem{remark}[theorem]{Remark}  
  
\newcommand{\Hmm}[1]{\leavevmode{\marginpar{\tiny%
$\hbox to 0mm{\hspace*{-0.5mm}$\leftarrow$\hss}%
\vcenter{\vrule depth 0.1mm height 0.1mm width \the\marginparwidth}%
\hbox to 0mm{\hss$\rightarrow$\hspace*{-0.5mm}}$\\\relax\raggedright #1}}}  
\begin{document}  
\title[Eigenfunction expansion]{Eigenfunction expansions for Schr\"odinger operators on metric graphs}  
\author[D.~Lenz]{Daniel Lenz$^{+,*}$}   
\author[C.~Schubert]{Carsten  
Schubert$^*$}  
\author[P.~Stollmann]{Peter  
Stollmann$^*$}   
\address{$^*$ Fakult\"at f\"ur  
Mathematik, Technische Universit\"at, 09107 Chemnitz, Germany\\
$^+$ Current address: Dept. of Mathematics, Rice University, P. O. Box 1892, Houston, TX 77251, USA}  
\begin{abstract}  
We construct an expansion in generalized eigenfunctions for   
Schrödinger operators on metric graphs. We require rather minimal   
assumptions concerning the graph  structure and the boundary conditions at the vertices.  
\end{abstract}  
\date{\today} %
\maketitle  

\section*{Introduction}  
Expansion in generalized eigenfunctions is a topic that dates back to
\textsc{Fourier's} work, at least. A classical reference is Berezanskii's
monograph \cite{Berezanskii-68}. Motivated by examples from Mathematical
Physics there has been a steady development involving new models. One trigger
of more recent results is the importance of generalized eigenfunction
expansions in the discussion of random models.  See
\cite{BoutetdeMonvelS-03b,KleinKS-02,Stollmann-01} and the references in
there.   This was also the background of the first paper that established
eigenfunction expansions for quantum graphs, \cite{HislopP}   (see   \cite{AizenmanSW-06b,HislopP,Kuchment-02,  
Kuchment-04,Kuchment-05,Exner-97,KostrykinS-99b,KostrykinS-00b,KostrykinS-06}
for recent results on quantum graphs).   There the authors
consider a rather special class of metric graphs, due to the random model they
have in mind.  We point out, however that part of their discussion is rather
abstract and pretty much equivalent to what had been obtained in
\cite{BoutetdeMonvelS-03b}.  As was pointed out in \cite{BoutetdeMonvelLS}, the
Dirichlet form framework of the latter article applies to a class of quantum
graphs with Kirchhoff boundary conditions. 
  
The point of the present paper is to establish an expansion in generalized
eigenfunctions under somewhat minimal conditions.  This means we require just
the usual conditions necessary to define the operators in question.  These
conditions essentially amount to providing a continuous embedding from the
form domain of the operator to the Sobolev space $W^{1,2}$ of the graph.   More
concretely, we allow for general boundary conditions, unboundedness of the
(locally finite) vertex degree function, loops, multiple edges and edges of
infinite lengths.  However, we require a uniform lower bound on the length of
the edges.   To the best of our knowledge, this framework contains all classes
of models that have been considered so far.   Our discussion is intrinsic and
does not require an embedding of the metric graph into an ambient space. 
  
As far as methods are concerned, we rely on the results from
\cite{PoerschkeSW-89} rather than the approach of \cite{Berezanskii-68} that
had been used in \cite{HislopP}.  However, this is mostly a question of habit. 
In either approach a main point is to establish certain trace class properties
of auxiliary functions.  Here, we can rely upon one-dimensional techniques for
quantum graphs.  An extra asset is that we are able to establish pointwise
properties of generalized eigenfunctions.

Our paper is structured as follows:
In Section 1 we set up model and notation, define metric graphs and introduce
the kind of boundary conditions we allow.  Moreover, we check the necessary
operator theoretic input for the Poerschke-Stolz-Weidmann method for
constructing generalized eigenfunctions. 
  In Section 2 we discuss the notion of generalized eigenfunctions and explore
pointwise properties in the quantum graph case.  It turns out that in this case
generalized eigenfunctions have versions that satisfy the boundary conditions
at the vertices.  In Section 3 we present the necessary material from
\cite{PoerschkeSW-89}.  The application to the quantum graph case comes in
Section 4 that contains our main results, Theorem \ref{mein_theorem} and
Corollary \ref{potential}.  The former deals with quantum graphs and the latter includes
additional perturbations by a potential that is uniformly locally square
integrable.

\section{Metric graphs and the associated operators}  
  
In this section we introduce metric graphs and the associated operators.    
The basic idea is that a metric graph consists of line  
segments -- edges -- that are glued together at vertices.  In contrast  
to combinatorial graphs, these line segments are taken seriously as   
differential structures  
and  
in fact one is interested in the Laplacian on the union of the line  
segments.  To get a self adjoint operator one has to specify boundary  
conditions at the vertices.  Our discussion of the unperturbed operator  
associated to a quantum graph in this section relies on the cited  
works of Kostrykin \& Schrader, Kuchment and the second named author.

\smallskip

\begin{definition}  
A \emph{metric graph} is $\Gamma=(E,V,i,j)$ where  
\begin{itemize}  
\item $E$ (edges) is a countable family of open intervals $(0,l(e))$  
and $V$ (vertices) is a countable set.   
\item $i:E\to V$ defines the initial point of an edge and $j:\{ e\in  
E| l(e)<\infty \}\to V$ the end point for edges of finite length.   
\end{itemize}  
We let $X_e:= \{ e\}\times e$, $X=X_\Gamma=V\cup\bigcup_{e\in E}X_e$  
and $\overline{X_e}:=X_e\cup\{ i(e),j(e)\}$.
\end{definition}  
Note that $X_e$ is basically just the interval $(0,l(e))$, the first  
component is added to force the $X_e$'s to be mutually disjoint.  The  
topology on $X$ will be such that the mapping $\pi_e:X_e\to (0,l(e)),  
(e,t)\mapsto t$ extends to a homeomorphism again denoted by  
$\pi_e:\overline{X_e}\to \overline{(0,l(e))}$ that satisfies  
$\pi_e(i(e))=0$ and $\pi_e(j(e))=l(e)$ (the latter in case that  
$l(e)<\infty$). 
A piece  of the form $I= \pi_e^{-1} (J)$ with an edge $e$ and  an interval $J\subset \overline{(0, l(e))}$  is called an \textit{edge segment}. The length of the edge segment is the length of $J$. 
Edge segments will play a role, when we discuss local properties of functions.

While we allow multiple edges and loops, we will assume  
finiteness of each single vertex degree $d_v$, $v\in V$, i. e.   
\begin{itemize}  
\item[(F)] $d_v:= |\{ e\in E |v\in\{ i(e),j(e)\}| <\infty$.   
\end{itemize}  
  
To define a metric structure on $X$ we then proceed as  
follows: we say that $p\in X^N$ is a \emph{good polygon} if, for every  
$k\in\{ 1, . . .  ,N\}$ there is a unique edge $e\in E$ such that  
$\{x_k,x_{k+1}\}\subset \overline{X_e}$. Using the usual distance in  
$[0,l(e)]$ we get a distance $d$ on $\overline{X_e}$ and define  
$$l(p)=\sum_{k=1}^Nd(x_k,x_{k+1}) . $$  
Since multiple edges are allowed, we needed to restrict  
our attention to good polygons to exclude the case that  
$\{x_k,x_{k+1}\}$ are joined by edges of different length. Given  
connectedness of the graph and (F),   
a metric on $X$ is given by  
$$  
d(x,y):=\inf\{ l(p)| p\mbox{ a good polygon with } x_0=x\mbox{ and  
}x_N=y\} .   
$$  
In fact, symmetry and triangle inequality are evident and the  
separation of points follows from the finiteness.  Clearly, with the  
topology induced by that metric, $X$ is a locally compact, separable  
metric space. If $X$ is not connected we can do the above procedure on any  
connected component.   
  
We will assume a lower bound on the length of the  
edges:  
  
\begin{itemize}  
\item[(LB)] There exists a $u>0$ with $l(e) \geq u$ for all $e\in E$.    
\end{itemize}   
  
We will now turn to the relevant Hilbert spaces and operators.    
We define   
\[  
L^2(X):= \bigoplus_{e\in E } L^2(e),\; \: W^{1,2} (X) := \bigoplus_{e\in E }
W^{1,2}(e), W^{2,2}(X):= \bigoplus_{e\in E } W^{2,2}(e).
\]  
Here, of course, $L^2(e) $ ($W^{1,2} (e)$, $W^{2,2}(e)$) consists of functions
$u_e$ on $e=(0,l(e))$. In the sequel we will view those families $u=(u_e)_{e\in E}\in L^2 (X)$
rather as functions defined on $X$.

Consider $a>0$ and recall that $h\in W^{1,2}(0,a)$ is continuous and
$h(0):=\lim_{x\to 0+} h(x)$ exists and satisfies
\begin{equation}\label{Sobolev}  
|h(0)|^2 \leq \frac{2}{a} \|h\|^2_{L^2 (0,a)} + a \|h'\|^2_{L^2 (0,a)}  
\end{equation}  
by standard Sobolev type theorems. Consider now an edge $e$, the vertex
$v=i(e)\in V$and $u\in W^{1,2} (e)$.  Then the limit $ u (v) :=\lim_{t\to 0} u
(t) $ exists, as well as $u (w) :=\lim_{t\to l(e)} u (t) $ for $w=j(e)$ and
\eqref{Sobolev} holds (with the obvious modifications).  Similarly, for an
edge $e$ and the vertex $v= i(e) $ and the vertex $w=j(e)$ and $u\in W^{2,2}
(e)$ the limits $ u' (v) :=\lim_{x\to v, x\in e} u' (x) \; \: u' (w)
:=-\lim_{x\to w, x\in e} u' (x)$ exist.  Here, we have introduced a sign.
This makes our definition of the derivative canonical, i. e. independent of
the choice of orientation of the edge.  For$f\in W^{1,2}(X)$ and each vertex
$v$ we gather the boundary values of $f_e (v)$ over all edges $e$ adjacent to
$v$ in a vector $f(v)$.  More precisely, let $E_v:= \{ e\in E| v\in\{
i(e),j(e)\}\}$ denote the set of vertices adjacent to $v$ and define $f(v):=
(f_e(v))_{e\in E_v}\in \CC^{E_v}$ and similarly, for $f\in W^{2,2}(X)$ we
further gather the boundary values of $f_e' (v)$ over all edges $e$ adjacent
to $v$ in a vector $f' (v)\in \CC^{E_v}$.  These boundary values of functions
will play a crucial role when discussing the concept of boundary condition.

\begin{definition}  
A \emph{boundary condition} is given by a pair $(L,P)$ consisting of a  
family $L= (L_v)_{v\in V}$ of self-adjoint operators $L_v :\CC^{E_v}  
\longrightarrow \CC^{E_v}$ and a family $P=(P_v)$ of projections $P_v  
:\CC^{E_v} \longrightarrow \CC^{E_v}$.   
\end{definition}  
We will assume the following upper bound on $(L_v)_{v\in V}$:  
  
\begin{itemize}  
\item[(UB)] There exists an $S>0$ with $\|L^+_v\| \leq S$ for any $v\in V$,  
where the $+$ denotes the positive part of a self-adjoint operator.   
\end{itemize}  
  
Given a metric graph satisfying (F) and(LB) and a boundary condition
satisfying (UB), we obtain from \eqref{Sobolev} by a direct calculation that
\begin{equation}\label{Randterm}  
\sum_{v\in V} \langle L_v f(v), f(v)\rangle \leq \frac{4 S}{\varepsilon} \|f\|_{L^2(X)} + 2 S \varepsilon\|f'\|_{L^2(X)}  
\end{equation}  
for any $f\in W^{1,2} (X)$ and any $\varepsilon >0  
$ with $ \varepsilon \leq u$.    
Given a boundary condition $(L,P)$ we define the form $s_0:= s_{L,P}$ by  
$$ D(s_0) :=\{f\in W^{1,2}(X) :P_v f(v) = 0\;\mbox{for all $v\in V$ } \} ,$$  
$$ s_0 (f,g) :=\sum_{e\in E} \int_0^{l(e)} f_e' (t)\overline{g}_e'(t) dt- \sum_{v\in V} \langle L_v f(v), \overline{g}(v)\rangle. $$  
  
By \eqref{Randterm} we easily see that for $C>0$ large enough  
  
\begin{equation}\label{Heins}  
s_0 (f,f) + C (f,f) \geq \frac{1}{2} \|f\|_{W^{1,2} (X)}  
\end{equation}  
for any $f\in D (s_0)$. This shows that $s_0$ is bounded below and closed.
Hence, there exists an associated self-adjoint operator.  This operator is
denoted by $H_0:=H_{L,P}$.  It can be explicitly characterized by
$$ D(H_0) := \{f\in W^{2,2} (X) : P_v f(v) =0, $$  
$$ L_v f(v) + (1 - P_v) f' (v)  
=0\;\:\mbox{for all $v\in V$ } \},$$  
$$ (H_0 f)_e:= - f_e'' \;\:\mbox{for all $e\in E$}. $$  
  
We will assume the following setting:  
  
\begin{itemize}  
\item[(S)] $\Gamma$ is a metric graph satisfying (F) and (LB) with associated  
space $X$.  $(L,P)$ is a boundary condition satisfying (UP).  The induced  
form is denoted as $s_0$ and the corresponding operator by $H_0=H_{L,P}$.    
\end{itemize}

\begin{lemma}\label{lemmaeins} Assume (S).    
Then, $(H_0 +C)^{-\frac{1}{2}}$ provides a
continuous map from $L^2 (X)$ to $L^\infty (X)$ for sufficiently large $C>0$.   
\end{lemma}  
\begin{proof}  
As $H_0$ is bounded below,$(H_0 +C)^{-\frac{1}{2}}$ provides a  
bounded map from $L^2 (X)$ to the form domain equipped with the form  
norm $\|\cdot\|_{s_0}$ for sufficiently large $C>0$.  By \eqref{Heins},   
 the form domain (with the form norm) is continuously embedded into  
$W^{1,2}(X)$.  By \eqref{Sobolev}, $W^{1,2}(X)$ is continuously embedded in  
$L^\infty (X)$.  Putting this together we obtain the statement.    
\end{proof}  
  
\begin{lemma}\label{lemmazwei} Assume (S). Then,  
$$\{ f\in D(H_0) : \supp f\;\mbox{compact}\}$$  
is a core for $H_0$.    
\end{lemma}  
\begin{proof}Choose $f\in D(H_0)$.  We have to find $f_n\in D(H_0)$ with  
compact support and $f_n \longrightarrow f$ and $H_0 f_n \longrightarrow H_0  
f$. We will provide $f_n = \psi_n f$ with suitable cut-off functions  
$\psi_n$.  We will assume without loss of generality that $X$ is connected (otherwise we will have to  
perform the process  simultaneously on each connected component).    
  
Choose $x\in X$.  For $n\in \N$ let $B_n=B(x,n)$ be the ball around $x$ with
radius $n$. Construct $\psi_n = (\psi_{n,e})_{e\in E}$ by distinguishing three
cases: For edges $e$ with both ends $i(e)$ and $j(e)$ contained in $B_n$ set $
\psi_{n,e}\equiv 1$.  For edges $e$with both ends $i(e)$ and $j(e)$ contained
in the complement of $B_n$ set $\psi_{n,e}\equiv 0$.  For edges $e$ with one
endpoint, say $i(e)\in B_n$ and $j(e)\in B_n^c$ we choose $\psi_{n,e}$ two
times continuously differentiable on $e$, $\psi_{n,e}\equiv 1$ on a
neighborhood of $i(e)$, $\psi_{n,e}\equiv 0$ on a neighborhood of $j(e)$ such
that $\psi_{n,e}$ and its first two derivatives are bounded by $ (1 +4/u)^2$.
This is possible, since the length of the edges is bounded below by $u$.
Since in this way $\psi_n$ is constant in the neighborhood of any vertex,
smooth and bounded the functions $f_n := \psi_n f\in D(H_0)$ for every $n\in
\NN$.  Clearly,
$$  
\psi_n |_{B(x;n-2u)} \equiv 1, \quad \psi_n |_{B(x;n+2u)^c}\equiv 0 ,  
$$  
so that $f_n\to f$ in $L^2(X)$ as $n\to\infty$.  Similarly,  
$$  
H_0( \psi_n f)=-\psi_n f''-2\psi_n' f'-\psi_n'' f\to H_0f ,  
$$  
as $\psi_n',\psi_n''$ are uniformly bounded and supported on $B(x;n+2u)\setminus B(x;n-2u)$.   
\end{proof}


\begin{remark} {\rm  Let us shortly discuss the necessity of  conditions of the form
  (LB) and (UB) in our context. Our aim is to show  \eqref{Heins}, i.e. that the identity is continuous as a map from the form domain with $W^{1,2}$ norm to the form domain with form norm.

As we allow for rather general boundary conditions and do not assume any connectedness, we need a pointwise estimate on the boundary values of a function on an edge  in terms of the corresponding $W^{1,2} (e)$ norm.  In this respect, the Sobolev estimate
  \eqref{Sobolev} is essentially optimal.  More precisely, testing with the
  constant function on an interval of finite length shows that the factor
  $1/a$ can not be avoided.  In particular, \eqref{Heins} fails for a graph consisting of countably infinite  disjoint edges with lengths going to zero and   a $\delta$ - boundary condition  (corresponding to  $L_v$ being the identity) on one of the vertices of each edge.   In this sense, a condition of the form (LB) seems unavoidable. 

Similarly, given (LB), we need a bound of the form (UB) to bound the boundary terms. In particular, \eqref{Heins} fails for a graph consisting of countably infinite  disjoint edges with lengths one and   boundary conditions of the form   $ c_v L_v$ with $c_v$ going to infinity. }
\end{remark}  

\section{A word on locality}\label{Local}  
Let a locally compact space $X$ with a measure $d x$ be given.  Let
$\eL^2_{\mathrm{loc}} (X)$ be the space of functions on $X$ whose restrictions
to compact sets are square integrable.  Let $\eL^2_{\mathrm{comp}} (X)$ be the
set of functions in $\eL^2 (X)$ which have compact support.  The usual inner
product can be ``extended'' to give a map (again denoted by
$\langle\cdot,\cdot\rangle$)
$$ \eL^2_{\mathrm{comp}} (X)\times \eL^2_{\mathrm{loc}} (X)\longrightarrow \CC,  
\langle f,g\rangle :=\int f (x) \overline{g} (x) dx. $$

\begin{definition} Let $X$ be a topological space with a measure $d x$.  Let  
  $H$ be an operator on $X$ which is local i. e.  $H f$ has compact support
  whenever $f$ has and $D(H)\cap \eL_{\mathrm{comp}}$ is a core for $H$.  A
  nontrivial function $\phi$ on $X$ is called a \emph{generalized eigenfunction} for
  $H$ corresponding to $ \lambda$, if it belongs to $\eL^2_{\mathrm{loc}} (X)$
  and satisfies
\begin{equation}\label{eq_genef}\langle H f, \phi\rangle = \lambda \langle f,  
\phi \rangle  
\end{equation}  
for any $f\in D(H)$ with compact support.    
\end{definition}  
\begin{remark}{\rm Here, $\langle H f, \phi\rangle$ is defined in the sense discussed  
at the beginning of the section.  The inner product $\langle f, \phi \rangle$ is  
defined in the same way.  The condition on the core of $H$  
is not necessary to  
state the definition.  However, it is only this condition that makes the  
definition a sensible one. }  
\end{remark}

The question arises to which extent a generalized eigenfunction is locally a  
good function.  We say that $\phi\in\eL^2_{\mathrm{loc}} (X)$ \emph{is locally in} $W^{2,2}$ if  the restriction $\phi_I$  belongs to $W^{2,2} (I)$ for any compact edge segment.   In particular, $\phi_e\in W^{2,2}(e)$ for every   
edge of finite length.  Note that $\eL^2_{\mathrm{loc}} (X)$-functions belong to $\eL^2$ of any edge of finite length.

\smallskip  
  
Here is one answer in the case of quantum graphs:

\begin{lemma} \label{genef}Assume (S).  If $\phi$ is a generalized eigenfunction for $H_0$, then $\phi$ is  
locally in $W^{2,2}$ and admits a version that satisfies  
the boundary condition at any vertex.    
\end{lemma}  
\begin{proof} To check that $\phi$ belongs locally to $W^{2,2}$ is suffices to consider $f\in D(H_0)$ with  
compact support contained in an edge and apply \eqref{eq_genef}.  This gives $-\phi'' =\lambda \phi$ so that   
$\phi$ belongs locally to $W^{2,2}$, since $\phi\in\eL^2_{\mathrm{loc}} (X)$ by our definition of   
generalized eigenfunction.   
   
To check that  
$\phi$ satisfies the boundary condition at a vertex $v$, it suffices to consider $f\in D(H_0)$  
supported on a neighborhood of $v$ and apply $\eqref{eq_genef}$.    
In fact, let $f\in D(H_0)$ with $f_e\equiv 0$ for $e\not\in E_v$.  Then we get  
\begin{eqnarray*}  
\langle f, \lambda \phi\rangle &=& \langle H_0 f, \phi\rangle \\  
 &=& \langle -f'', \phi\rangle   
\end{eqnarray*}  
integration by parts and the condition on the support of $f$ give (with the evident notation for the inner product in   
$\CC^{E_v}$):  
  
\begin{eqnarray*}  
. . .  &=& \langle f, -\phi''\rangle +\langle f'(v),\phi(v) \rangle - \langle f(v),\phi'(v) \rangle\\  
&=& \langle f, \lambda\phi\rangle +\langle f'(v),\phi(v) \rangle - \langle f(v),\phi'(v) \rangle  
\end{eqnarray*}  
as the second weak derivative of $\overline{\phi}$ is
$-\lambda\overline{\phi}$.  Therefore,
$$  
\langle f'(v),\phi(v) \rangle = \langle f(v),\phi'(v) \rangle  
$$  
for every choice of $f\in D(H_0)$.  Splitting the scalar products in the parts living in the images of $P_v$ and $(1-P_v)$  
gives:  
  
\begin{align*}  
\langle P_v f'(v),P_v \phi(v) \rangle&+\langle (1-P_v) f'(v),(1-P_v) \phi(v) \rangle \\  
= \langle P_v f(v),P_v \phi'(v) \rangle&+\langle (1-P_v) f(v),(1-P_v) \phi'(v) \rangle.  
\end{align*}  
Choosing $f\in D(H_0)$ with arbitrary $P_v f'(v)$ and $(1-P_v)f(v)=0$ (granting $(1-P_v)f'(v)=0$), we see that   
$P_v\phi(v)$ has to be equal to zero.   
  
If we use the boundary conditions for $f$ the last equation can be transformed to   
$$  
\langle P_vf'(v),P_v\phi(v) \rangle =\langle (1-P_v)f(v) , L_v\phi(v)+(1-P_v)\phi'(v) \rangle.   
$$  
 Taking an $f$ with arbitrary $(1-P_v)f(v)$, we conclude that   
$L_v\phi(v)+(1-P_v)\phi'(v)$ also equals zero, thus giving the boundary conditions for $\phi$.   
\end{proof}

\section{Expansion in generalized eigenfunction: General framework}  
  
In this section we discuss the expansion in generalized eigenfunctions of a
self-adjoint operator.  We follow the work of Poerschke, Stolz and Weidmann
\cite{PoerschkeSW-89}.  This will be used to provide an expansion for metric
graphs in a spirit similar to the considerations of \cite{BoutetdeMonvelS-03b}
for Dirichlet forms.  Note that in \cite{HislopP} a different approach has
been used.  However, an important point in both the different methods is to
establish suitable trace class properties for operators constructed from $H$.
In that respect, the analysis of \cite{BoutetdeMonvelS-03b,HislopP} is
similar.  Actually, the case of quantum graphs is rather easy as far as trace
class properties are concerned, as we have a locally one-dimensional situation
at hand.  \smallskip

Let a Hilbert space $(\hil,\langle \cdot , \cdot \rangle)$ and a self-adjoint
operator $T\geq 1$ in $\hil$ be given.  We will define the following two
auxiliary Hilbert spaces: $\hil_+:= \hil_+(T) :=D(T)$, $\langle x,y
\rangle_+:=\langle Tx,Ty \rangle$ and $\hil_-$ as completion of $\hil$ with
respect to the scalar product $\langle x,y\rangle_-:=\langle T^{-1}x,T^{-1}y
\rangle$.  Thus, the inner product on $\hil$ can be naturally extended to give
a map
$$ \langle \cdot, \cdot \rangle : \hil_+\times \hil_- \longrightarrow \CC. $$  
Let $N$ be a positive integer or infinity, $H$ a self-adjoint operator in $\hil$ and $\mu$ a spectral measure for $H$.

A sequence of subsets $M_j\subset \RR$, such that $M_j \supset  
M_{j+1}$ together with a unitary map $U$  
\begin{equation*}  
 U=(U_j):\hil\to \oplus_{j=1}^N L^2(M_j,d\mu)  
\end{equation*}  
is said to be an \textit{ordered spectral representation} of $H$ if  
\begin{equation*}  
 U\phi(H)=M_\phi U,  
\end{equation*}  
for every measurable function $\phi$ on $\RR$.   
  
\begin{theorem}  
Let $H$, $T$, $\hil_+$, $\hil_-$ be as above.  Let $\mu$ be a spectral measure  
for $H$ and $U$ an ordered spectral representation. Let $\gamma : \RR  
\longrightarrow \CC$ be continuous and bounded with $|\gamma|>0$ on $\sigma(H)$   
such that $\gamma(H)T^{-1}$ is a Hilbert-Schmidt operator.    
Then there are measurable functions $\phi_{j}:M_j\to\hil_-, \lambda\mapsto \phi_{j,\lambda}$ for  
$j=1,\ldots,N$ such that the following properties hold:  
\begin{itemize}  
\item[(i)] $  
U_jf(\lambda)=\langle f,\phi_{j}(\lambda)\rangle \text{ for } f\in \hil_+ \text{ and }\mu\text{-a. e.  }\lambda \in M_j.  $  
\item[(ii)] For every $g=(g_j)\in \oplus_j L^2(M_j,d\mu)$   
\begin{equation*}  
Ug=\lim_{n\to N, E\to \infty} \sum\limits_{j=1}^N \int\limits_{M_j\cap [-E,E]}g_j(\lambda)\phi_{j,\lambda}d\mu(\lambda)  
\end{equation*}  
and, for every $f\in \hil$,  
\begin{equation*}  
f=\lim_{n\to N, E\to \infty} \sum\limits_{j=1}^N \int\limits_{M_j\cap [-E,E]} (U_jf)(\lambda)d\mu(\lambda).   
\end{equation*}  
\item[(iii)]For $f\in \{ g\in D(H)\cap\hil_+ | \ Hg\in \hil_+\}$   
\begin{equation}  
\label{eq_genefexp}  
\langle Hf,\phi_{j,\lambda} \rangle =\lambda \langle f,\phi_{j,\lambda} \rangle.   
\end{equation}  
\end{itemize}  
\end{theorem}  
  
If the functions $\phi_{j,\lambda}$ fulfill (i) and (ii) of the theorem we  
will speak of a Fourier type expansion.   
 If the set $\{ g\in D(H)\cap\hil_+ | \ Hg\in \hil_+\}$ is a core for $H$ we  
 speak of an expansion in generalized eigenfunctions.

\smallskip  
  
We will apply the previous Theorem to the Hilbert Space $\hil=L^2(X)$ and the
operator $H_0$, where $X$ is a quantum graph satisfying (F), (LB) and (UB) as
discussed in Section 1.  As operator $T$ we define a multiplication operator
$T:=M_w$ as multiplication with a weight function $w$, i. e.  a continuous map
$w : X \longrightarrow [1,\infty)$.

\section{The main theorem}

\begin{theorem}  
\label{mein_theorem}Assume (S).   
Let $\mu$ be a spectral measure for $H_0$.  Let $w:X\to [1,\infty)$ be
continuous with $w^{-1}\in L^2(X)$.  Then there exists a Fourier type
expansion $(\phi_j)$ for $H_0$, such that for $\mu$-a. e.  $\lambda \in
\sigma(H_0)$ the function $\phi_{j,\lambda}$ is a generalized eigenfunction of
$H_0$ for $\lambda$ with $w^{-1}\phi_{j,\lambda}\in L^2$.
\end{theorem}

\begin{proof} We will apply the abstract result of the previous section.  Let  
$\gamma$ be the function $\gamma (t) = ( C + t)^{-1/2}$.   
As $T$ choose multiplication with $w$.  Then,  
$\gamma(H_0)$ is a bounded map from $\eL^2 (X)$ to $\eL^\infty (X)$ by Lemma  
\ref{lemmaeins}.  This, together with the assumption on $w$ easily shows  
that the operator $T^{-1} \gamma(H_0)$ has an $\eL^2$ kernel and is therefore  
a Hilbert-Schmidt operator.  Thus, its adjoint operator $\gamma (H_0) T^{-1}$  
is a Hilbert-Schmidt operator as well.  We can therefore apply the result of  
the previous section.  This gives a Fourier type  expansion.  By definition of $T$ any function in $\hil_-$ is locally in $L^2$. 

Moreover, 
\begin{equation*}  
\langle H_0 f,\phi_{j,\lambda} \rangle =\lambda \langle f,\phi_{j,\lambda} \rangle   
\end{equation*}  
holds $\mu$-a. e.  for $f \in D_w :=\{g \in D(H_0)| \ wg, wH_0 g \in
L^2(X)\}$.  As $w$ is continuous and $H_0 f$ has compact support whenever $f$
has compact support by definition of $H_0$, the set $D_w$ obviously contains
$D(H_0) \cap \eL^2_{\mbox{comp}} (X)$.  Thus, the functions $\phi_{j,\lambda}$
are generalized eigenfunctions in the sense of Section \ref{Local}.  This
finishes the proof.
\end{proof}  
  
We denote by $m$ the measure induced on $X$ by the Lebesgue measure on the edges $X_e$, pulled back via $\pi_e$.   
\begin{remark}  
 {\rm (\textbf{A weight function.}) Assume that $X$ is connected and define, for $\epsilon >0$,  
\begin{equation*}  
 w(x)=m\left( B_{d(x,x_0)+1}(x_0)\right)^{1+\epsilon}.   
\end{equation*}  
Clearly, $w$ is continuous and $w\geq 1$.  To see that $w^{-1}\in L^2(X)$, it
suffices to consider the case that $\Gamma$ is infinite.  In this case,
$m(B_r(x_0))\ge r$ for every $x_0\in X$ and $r>0$ by construction of the
metric.  We consider the volume of the annuli $B_n(x_0)\setminus
B_{n-1}(x_0)$.  For $x$ in this annulus we obviously have that $w(x)\geq
m(B_{n}(x_0))^{1+\epsilon}$.
\begin{align*}  
  \int\limits_X |w^{-1}|^2 dx &\leq \int\limits_{B_1\setminus B_0} w^{-2}dx+ \int\limits_{B_2\setminus B_1} w^{-2}dx+\ldots
\end{align*}
\begin{align*}
  &\leq \int\limits_{B_1\setminus B_0}m(B_1)^{-2-2\epsilon}dx+ \int\limits_{B_2\setminus B_1} m(B_2)^{-2-2\epsilon}dx+\ldots\\
  &\leq \sum_{i=1}^\infty i^{-1-2\epsilon}<\infty ,
\end{align*}  
where we used $m(B_n)\geq n$. }  
\end{remark}  
\subsection*{Schr\"odinger operators}  
  
Now we show that our main result can be extended to Schr\"odinger operators on
metric graphs.  Here, we treat a rather simple case.  More singular
perturbations will be considered elsewhere.  In the following proposition we
gather some operator theoretic results for potential perturbations of the
operators $H_0=H_{L,P}$ for a quantum graph satisfying assumption (S).  For a
general background, we refer the reader to \cite{ReedS-75}, Section X. 2 as
well as \cite{Faris-75}, § 5 and § 6.
  
We are going to consider the class  of potentials $V   \in \prod\limits_e\eL^2(e)$ with
$$  
M:=M_V:=\sup \{\| V_I \|_2 : I \;\mbox{edge segment with length between $u$ and $2 u$} \} <\infty .   
$$  
This class will be denoted by $\eL^2_{\mathrm{loc,u}}(X)$.

  
\begin{prop}  
Assume (S) and let $V\in \eL^2_{\mathrm{loc,u}}(X)$.  Then we have:  
\begin{itemize}  
\item[\textrm{(i)}] $V$ is infinitesimally small with respect to $H_0$.   
In particular, $H=H_0+V$ is self-adjoint on $D(H_0)$.   
\item[\textrm{(ii)}] $(H_0 +C)^{-\frac{1}{2}}$ provides a  
continuous map from $L^2 (X)$ to $L^\infty (X)$ for sufficiently large $C>0$.   
\item[\textrm{(iii)}] $$\{ f\in D(H_0) : \supp f\;\mbox{compact}\}$$  
is a core for $H$.   
\item[\textrm{(iv)}]If $\phi$ is a generalized eigenfunction for $H$, then $\phi$ is  
locally in $W^{2,2}$ and admits a version that satisfies the boundary condition at any vertex.    
\end{itemize}  
\end{prop}  
\begin{proof}  
(i) Let $a>0$ be arbitrary.  Assume w.l.o.g that $a\leq u$.  We now decompose the  edges of the graph into edge segments, which are disjoint up to their boundary and have length between $a$ and $2a$. 
  Then,  any point of the graph belongs to  such a edge  segment $I$.  Accordingly, our usual Sobolev estimate   \eqref{Sobolev} gives

\begin{equation}\label{usual}
 \|f|_I\|_\infty^2 \leq \frac{a}{2} \| f'|_I\|_2^2 + \frac{4}{a} \|f|_I\|_2^2.
\end{equation}
Note that we pick up an extra factor of $2$ compared to  estimate   \eqref{Sobolev} as the point may not lie at the boundary of $I$ (in which case we only have an interval of length $a/2$ at our disposal).
Recall the estimate
\begin{equation}\label{three}
\|f\|_{W^{1,2}}^2 \leq  2 s_0 (f,f) +  C \|f\|_2^2.
\end{equation}
Summing over all $I$ of our decomposition we obtain
\begin{eqnarray*}
\|V f\|_2^2 & = & \sum_I \|  (V f)|_I\|_2^2\\
&\leq & \sum_I \| V|_I\|_2^2   \| f|_I\|_\infty^2\\
&\leq & M^2 \sum_I \| f|_I\|_\infty^2\\
\eqref{usual} &\leq &  M^2 \sum_I ( \frac{a}{2} \|f'|_I\|_2^2 +\frac{4}{a} \|f|_I\|_2^2)\\
&\leq & M^2 \frac{a}{2} \|f\|_{W^{1,2}}^2 +  M^2 \frac{4}{a} \|f\|_2^2\\
(\eqref{three}) \:\;\: & \leq & M^2 a s_0 (f,f) +  M^2 C \|f\|_2^2 +  M^2 \frac{4}{a} \|f\|_2^2\\
&=& M^2 a s_0 (f,f) + C(a) \|f\|_2^2,
\end{eqnarray*}
where 
$$ C(a) = M^2 (C +\frac{4}{a}).$$
As $s_0 (f,f) \leq \|H_0 f\| \|f\| \leq \|H_0\|^2 + \|f\|^2$, we obtain
$$\|V f\|^2 \leq  M^2 a \|H_0\|^2 + (C(a) + M^2 a)\|f\|^2_2.$$
As $a>0$ is arbitrary,  Self-adjointness of $H$ and (iii) both follow from the   
Kato-Rellich theorem, cf \cite{ReedS-75}, Theorem X. 12.   
  
(ii) It follows from (i) that $V$ is also form small with respect to $H_0$, see  
\cite{Faris-75} and \cite{ReedS-75} so that the form norm of $H_0$ and $H$ are equivalent. Hence (ii) follows from Lemma \ref{lemmaeins} above.   
  
(iv) For every compact edge segment $I$ we get that the restriction $\phi_I$ of $\phi$ to $I$ satisfies  
$$  
\phi_I''=V_I\phi_I-\lambda\phi_I 
$$  
in the weak sense.  Since $\phi_I\in\eL^2(I)$ for every compact $I$ and $V_I\in\eL^2(I)$, we  
get that $\phi_I\in L^1$.  In particular, $\phi_I'$ admits a continuous version so that  
$\phi_I\in C(I)$.  Since $V_I\in \eL^2(I)$ this gives that $\phi$ is locally in $W^{2,2}$.  The rest of the argument can be taken from the proof of Lemma \ref{genef} with the obvious rewording.    
\end{proof}  
  
This gives the following analog of Theorem \ref{mein_theorem} for Schr\"odinger operators:  
\begin{coro}\label{potential}  
Assume (S) and let $V\in\eL^2_{\mathrm{loc,u}}(X)$.   
Let $\mu$ be a spectral measure for $H$.  Let $w:X\to [1,\infty)$ be  
continuous with $w^{-1}\in L^2(X)$.  Then there exists a Fourier type expansion $(\phi_j)$ for $H$,   
such that for $\mu$-a. e.  $\lambda \in \sigma(H_0)$ the function $\phi_{j,\lambda}$ is a generalized   
eigenfunction of $H$ for $\lambda$ with $w^{-1}\phi_{j,\lambda}\in L^2$.   
\end{coro}

\end{document}